\documentclass[prd,aps,
tightenlines,
superscriptaddress,showpacs,nofootinbib
,eqsecnum,amsfonts,amsmath
]{revtex4}
\usepackage{bm,graphics,graphicx,epsfig,color
,subfigure,rotating,latexsym
}
\usepackage{amsthm}
\usepackage{amssymb}
\usepackage{url}
\usepackage{fancyhdr}
\usepackage{alltt}
\usepackage{caption}
\def\be{\begin{equation}}
\def\ee{\end{equation}}

\def\bea{\begin{eqnarray}}
\def\eea{\end{eqnarray}}

\begin{document}

\title[A stochastic template placement algorithm]
{A stochastic template placement algorithm for gravitational wave data analysis}

\author{I.\ W.\ Harry}
\email{ian.harry@astro.cf.ac.uk}
\affiliation{School of Physics and Astronomy, Cardiff University, Queens Buildings, The Parade, Cardiff, CF24 3AA, UK}
\author{B.\ Allen}
\email{Bruce.Allen@aei.mpg.de}
\affiliation{Albert Einstein Institute, Hannover, Germany and Department of Physics, U. Wisconsin - Milwaukee, Milwaukee WI USA}
\author{B.\ S.\ Sathyaprakash} 
\email{B.Sathyaprakash@astro.cf.ac.uk}
\affiliation{School of Physics and Astronomy, Cardiff University, Queens Buildings, The Parade, Cardiff, CF24 3AA, UK} 
\begin{abstract}
  This paper presents an algorithm for constructing matched-filter
  template banks in an arbitrary parameter space. The method places
  templates at random, then removes those which are ``too close''
  together.  The properties and optimality of stochastic template banks generated in
  this manner are investigated for some simple models. The
  effectiveness of these template banks for gravitational wave
  searches for binary inspiral waveforms is also examined.  The
  properties of a stochastic template bank are then compared to the
  deterministically placed template banks that are currently used in
  gravitational wave data analysis.
\end{abstract}
\maketitle

\section{Introduction}
Gravitational wave interferometric detectors have recently
completed a
science run in which a year of coincident data was taken at
the design sensitivity
\cite{Abbott:2009li,Acernese:2006bj}.
The data from this science run has been searched for gravitational
wave signals from compact binary inpsirals, unmodelled transient sources,
periodic sources and stochastic signals e.g.
\cite{abbott:122001,Abbott:2007ce,Abbott:2007wu}.
For many of these searches, analysis of the data from such
detectors takes advantage of the fact that the waveforms can be predicted
in advance and thus used in carrying out the analysis \cite{Th300}

The commonly used detection strategy to search for known signals in additive,
Gaussian, stationary noise
is the method of matched filtering
\cite{Helstrom68}. One correlates the (whitened) detector data with a
{\em template} (or {\em filter}), which is the (whitened) expected
signal waveform. The parameters of the source (sky position and
rotational frequency of a spinning neutron star, the masses of the
compact stars in a binary system, etc.)  are not known a-priori, so
the data must be correlated with many possible expected signal
waveforms, which have different values of parameters. The collection
of these points in parameter space is called a {\em template bank} or
{\em template grid}.

The past two decades have seen the development of methods
\cite{Schutz89,Sathyaprakash:1991mt,Dhurandhar:1992mw,Sathyaprakash:1994nj}
for
setting up template banks which minimize
the computational cost in a search without reducing the detectability
of signals. For instance, a geometric framework was 
developed \cite{BalSatDhu95,BalSatDhu96,Owen96}
in the 1990's to address the problem of template placement.
This
works quite well when the parameter space is of a small dimension (2, 3, or 4 at most)
\cite{OwenSathyaprakash98,BrCrCutSchu98,Bank06,Cokelaer:2007kx}.
The most important tool in this geometric framework is a
positive-definite {\em metric} which measures the fractional loss in
(squared) signal-to-noise ratio of a putative signal (at one
point in the parameter space) filtered through the optimal filter
corresponding to a nearby point in the parameter space.  The metric gives
the parameter space the geometric structure of a (possibly curved) Riemannian
manifold, which is often called the {\em signal manifold} (in this paper we
continue to refer to it as the parameter space).

When the dimension of the parameter space becomes large there are
problems with existing methods. First, even for flat parameter
spaces, there are no known optimal placement algorithms for dimensions
greater than 5 (the analogue of the two-dimensional hexagonal
lattice) \cite{Prix:2007rb} (and references therein).
Second, it is not clear how to place templates in a curved
parameter space. For example, one cannot set up an optimal (equally-spaced)
lattice on a two-sphere unless the number of points is very small (for example, 12).
This issue becomes increasingly important in parameter
spaces with dimension greater than 2. Third, if the parameter space
includes irregular boundaries, or is formed of regions with differing
dimensions, it is extremely difficult to ``step around'' the parameter
space in a deterministic way that covers the parameter space
completely but does not significantly over-cover it.

This paper gives a template placement algorithm that works for any
parametrized signal model in any number of dimensions, provided that
one can determine if two points in the parameter space are a large metric
distance apart, and, if they are not, accurately calculate the metric
distance between them.  The idea is simple.  Pick points at random in
parameter space, rejecting any points that are too close to those
previously retained.  Continue this process until no new points are
added, because any newly selected random points are close to
previously retained points.  We call this a {\em stochastic template
  placement} algorithm, and the resulting grid a {\em stochastic
  template grid} or {\em stochastic template bank}.

By construction, the stochastic template bank does not over-populate
the parameter space. But does it properly populate all regions?  The
answer depends upon the properties of the signal manifold and its
metric.  It is very similar to the question of whether the Monte-Carlo
approximation to an integral converges to the correct value.  And in
the same way as with Monte-Carlo integration, these stochastic
template banks appear to perform very well in real-world applications.

This method is closely related to another way of creating random
template banks, \cite{Messenger:2008ta}, in which the filtering stage
is not carried out, but has certain advantages.  In
particular, fewer templates are needed to obtain a given degree of
coverage of the parameter space. However, the filtering stage can
become computationally expensive.

Some practical issues remain.  The most convenient way to generate a
random template bank is to use computer-generated
uniformly-distributed random numbers as random coordinate values in
parameter space.  However, the distribution of the resulting points
then depends strongly upon the choice of the coordinate system.  If
global coordinates can be found in which the determinant of the metric
is constant (or nearly constant) then choosing uniformly distributed
random numbers for the coordinate values will result in a uniform
density of points.  This is optimal. If not, the random points should
ideally be generated with a probability density in coordinate space
proportional to the square root of the determinant of the metric in
those coordinates.  (One can also pick a small number of points in the
space, and at each point define a local coordinate system in which the
metric is proportional to $\delta_{ab}$, then place many points
uniformly in those coordinates.)  In practice, this is not necessary:
this paper shows that a stochastic template bank can still be
effectively generated by choosing uniform probability distributions
for the coordinate values, even if the determinant of the metric is
{\em not} constant on those coordinates. The only downside is
additional computational cost.

Later in this paper, two examples are shown to illustrate this: the
placement of templates in a $D$-dimension cube, and the placement of
templates on the signal manifold of gravitational wave chirps from
inspiralling compact binaries calculated in the first post-Newtonian
approximation.  In both cases, one can create stochastic template
banks using coordinates (polar, and masses $(m_1, m_2),$ respectively)
in which the determinant of the metric is not constant.  This incurs
unnecessary computation cost, but it works.  Alternatively, one can
create a stochastic template bank using coordinates (Cartesian, and
chirp-time coordinates $(\tau_1, \tau_3),$ respectively) in which the
metric (and hence its determinant) is approximately constant
\cite{Sathyaprakash:1994nj}.  This works better, since it is
computationally more efficient, but the end result is the same.

The paper is organized as follows. Sec.~\ref{sec:algorithm} presents
the stochastic template placement algorithm. An implementation and
results of testing are presented in Sec.~\ref{sec:tests} for some
simple cases where the number of templates is known
analytically. Sec.~\ref{sec:chirps} is devoted to the application of
the algorithm to the case of gravitational wave chirps from
inspiralling compact binaries where the performance of the stochastic
template placement method is compared with existing geometrical template placement
algorithms.

\section{Stochastic template placement algorithm}
\label{sec:algorithm}

Let $\cal M$ denote a signal manifold of dimension $D$, with $d(x,y)$
being a positive-definite distance function. Here $x,y \in \cal M$ are
points in the manifold.  Note that the signal manifold $\cal M$ might
cover only part of the space of
possible signals of a particular type, for example one might only want to
lay a bank to search for binary inspiral signals within a specific range
of masses.

A template bank $T$ is a set of $n$ points taken from $\cal M$:
$T=\{x_1, \cdots, x_n; \, x_i \in \cal M \}$.  A template bank is said to cover the signal
manifold with radius $\Delta$ (or to be complete) if every point in $\cal M$
lies within distance $\Delta$ of at least one of the $n$ points: $\forall y \in
{\cal M}, d(y, x_i) \le \Delta$ for at least one $i \in 1, \cdots, n$.

An optimal template bank of radius $\Delta$ would fulfill two conditions.  First, it would
cover the signal manifold with radius $\Delta$.  Second, it would
contain the minimum number of points.  However, it is difficult to
achieve this in practice!

The method proposed in this paper creates a template bank according to
the following algorithm:
\begin{itemize}
\item[1.] Let $T$ be a list of $n$ points from $\cal M$.  Initially
  $n=0$ and the list is empty.  As points get added to this list,
  they will be denoted by $x_1, \cdots, x_n$.
\item[2.] Pick a point $z$ at random from $\cal M$.  If $d(z,
  x_i)>\Delta$ for all points in the list $T$, then add $z$ to $T$ and
  increment $n$ by one.  Else discard the point $z$.
\item[3.] Repeat the previous step, until the list $T$ stops changing
  in length, or some other stopping criterion is met.
\end{itemize}

\subsection{Expected size of complete stochastic template banks}

An important question to ask is at what point will this iterative process
terminate?  This is
determined by the number of
templates needed to completely cover the space.  To
understand this, it is useful to first ask the more general question,
how large does a complete template bank (not necessarily one generated by the
algorithm above) need to be? To try to understand these questions this
sub-section begins by discussing upper
and lower bounds on the size of the stochastic
bank. Two commonly used lattice algorithms are then discussed and the 
performance of the stochastic bank,
at low dimension, is compared to these quantities.

In this discussion we follow \cite{Conway:1993,Messenger:2008ta} and
use
\emph{thickness} ($\Theta$) and \emph{normalized thickness} ($\theta$)
to assess the efficiency of a specific template covering.
Thickness is defined \cite{Conway:1993} as the average number of templates
covering any point in the parameter space while normalized thickness is defined as
the number of templates per unit volume in the case where
the radius of the templates is unity. They are related by \cite{Conway:1993}
\begin{equation}
\theta = \Theta/V_S.
\end{equation}
where $V_S$ is the volume enclosed by a D-dimensional sphere of unit radius
\begin{equation}
V_{S} = \frac{2 \pi^{D/2}}{ D \; \Gamma(D/2)}.
\end{equation}
The advantage of using
these quantities is that they are independant of the size of the parameter space
and independant of the template radius. These quantities are also directly related to the
number of templates that will be required \cite{Messenger:2008ta}, by
\begin{equation}
\theta = \frac{n \Delta^D}{V}
\end{equation}
where $V$ is the proper $D-$volume of the parameter space,
\begin{equation}
V = {\int_{\cal M} \sqrt{g}\, {\rm d}^D x}
\end{equation}
and $g$ is the determinant of the metric  $g_{ij}$ on the manifold $\cal M$.

We also assume, in this section, that ``boundary effects'' can be
ignored.  Except in pathological cases, this is true if the total
volume within distance $\Delta$ of $\partial \cal M$ is small compared
with the total volume of $\cal M$.

\begin{figure}
\includegraphics[width=3in,angle=0]{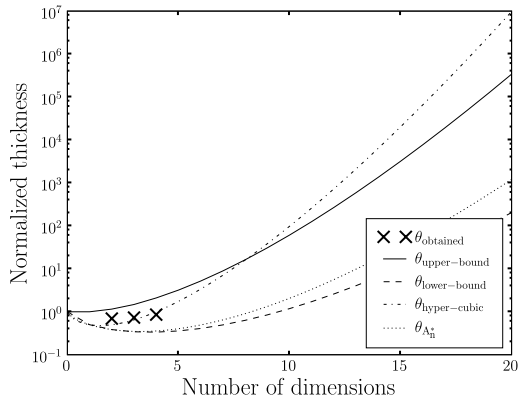}
\caption{\label{fig:comparison} The theoretical upper and lower
bounds on normalized thickness of a stochastic
template bank \cite{Conway:1993} and 
the normalized thickness of known lattice algorithms as a function
of dimension as defined by equations  (\ref{eq:upperbound}) and (\ref{eq:astar}).
Also the obtained thickness of
stochastic banks at dimensions less than 5.}
\end{figure}

A simple theoretical lower bound on the number of templates needed in any
complete template bank
is the ratio of the volume of the parameter space to the volume of a
single template.
The volume of a single template is the $D-$volume contained in a ball
of radius $\Delta$ is given by
\begin{equation}
V_{\rm template} = B(\Delta) = V_S \Delta^D.
\end{equation}
Hence the number of required templates is bounded below by
$V/V_{\rm template}.$
Alternatively we can say that the thickness of a complete template bank must be
greater than unity or that the normalized thickness must be greater than
$1/V_S$.
For the case of flat spaces a great deal of work has been carried out in trying
to obtain better estimates of the minimum possible thickness for a complete
template bank, it is clear, for example, that even in the 2 dimensional case a
complete template bank cannot have a thickness of 1, there must be some overlap
between the templates. In
\cite{Conway:1993} the best currently known theoretical
bounds on thickness are given and
these are the values that are shown as the lower
bound in Figure \ref{fig:comparison}.

\begin{figure}
\includegraphics[width=3in,angle=0]{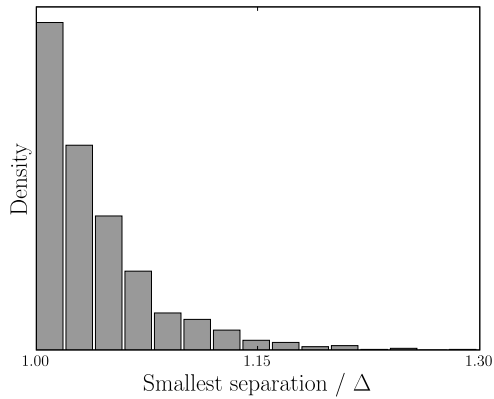}
\caption{\label{fig:distances} A histogram of the distribution of distances from a
  template to the nearest template, in units of the closest possible
  spacing $\Delta$, for a simple three-dimensional example. The
  distances are clustered close to the minimum possible spacing
  $\Delta$, showing that the covering locations found by the
  stochastic template placement method are close to the positions
  found by packing spheres of radius $\Delta/2$.}
\end{figure}

To try to obtain an upper bound on the thickness of the stochastic template
bank one can consider the sphere-packing problem, this is the question of how
many non-overlapping spheres can be packed into a certain volume.
Consider the packing problem with
hard spheres of radius $\Delta/2$. Since the centers of any of these
spheres are distances of $\Delta$ or more apart, they are suitable
locations for a stochastic template bank.   In Ref.\
\cite{Conway:1993} a bound is given on the number of hard
spheres of radius $\Delta/2$ that can be placed into a volume $V$.
This can be considered as an upper bound on the number of templates
that the stochastic algorithm can place. Figure \ref{fig:distances}
also suggests that this bound may be a reasonable estimate of the
thickness of a complete stochastic bank, it can be seen
that the average minimum distance between any template and the rest of the
bank is close to $\Delta$, as it would be in the sphere packing problem.
However, at least for low dimension ($D<4$), Figure~\ref{fig:comparison} shows that
the a complete stochastic template bank requires considerably fewer templates than this sphere
packing upper bound.

It is also useful to compare this with the performance of known lattice
algorithms. In this work two different lattice algorithms are considered.
The first is the hyper-cubical
lattice, where the hyper-cubes of the lattice are just small enough to fit
entirely inside a single ball of radius $\Delta$.  The side length
$\delta$ of such a cube is given by
\begin{equation}
\delta = 2 \Delta \left(1/D\right)^{0.5},
\end{equation}
since the longest diagonal of this $D$-cube is then $2 \Delta$ long.
Thus
\begin{equation}
\theta_{\rm hyper-cubic} = \frac{D^{D/2}}{2^D}
\label{eq:upperbound}
\end{equation}
describes the normalized thickness of a template bank in a
hyper-cubic arrangement.

The second lattice algorithm considered in this work is the $A_n^*$ lattice
\cite{Conway:1993,Prix:2007rb}. The 2 dimensional $A_n^*$ lattice is the well
known hexagonal lattice. For this algorithm the normalized thickness
is given by \cite{Messenger:2008ta}
\begin{equation}
\theta_{A_N^*} = \sqrt{D + 1} \left[
\frac{D\left(D+2\right)}{12\left(D+1\right)}\right]^{D/2}.
\label{eq:astar}
\end{equation}
From Figure \ref{fig:comparison} it can be seen that the $A_n^*$ lattice
requires less templates than the hyper-cubic lattice in all dimensions
(except the trivial one-dimensional case). It is also the most efficient
lattice known in dimensions up to 20 \cite{Conway:1993}.
This figure also shows that the
number of templates required to create a complete stochastic bank
is less than the hyper-cubic lattice, but only when the
dimension $D$ is greater than 3. A stochastic template bank with full
coverage, however, will require more templates than the $A_n^*$ lattice at least up
to four dimensions. We have no reason to believe that a complete stochastic bank will
be more efficient than the $A_n^*$ lattice in any dimension.

One must consider however that these lattice algorithms are only defined in the
case of flat parameter spaces. The stochastic algorithm on the other hand can be
used in any parameter space and it is in the cases where the parameter space is
not flat that we believe the stochastic bank would be the most useful.

\subsection{The convergence of a stochastic template bank}

In real world applications it may not be necessary for the template
bank to be complete. It is therefore useful to be able
to understand the convergence of the iteration that creates a
stochastic template bank. This sub-section is devoted to trying to understand
this convergence and comparing it
to the method describe in \cite{Messenger:2008ta}.

To begin to understand how a stochastic bank converges
it is necessary to define a \emph{covering
  fraction} $f \in [0,1]$.  The covering fraction is the ratio of the
volume of the subset of $\cal M$ that lies within a distance $\Delta$
of the points in the template bank, to the total volume of $\cal M$.
The expected number of trials required to add a new template to the
list is given by $1/(1-f)$, as can be seen by considering the template
placement process as a form of Monte-Carlo integration.

At the beginning of the iterative process, the template bank is empty,
and $f=0$.  After the first template is added (and assuming that
boundary effects can be ignored!) the covering fraction is
$f=\epsilon$, where $\epsilon = V_{\rm{template}}/V$, which is the
fraction of the entire volume covered by a single template. 
During the first iterative steps, while the number
of templates $n$ in the bank is small, $n \ll  1/\epsilon$, the
covering fraction increases linearly with the template number
according to $f = \epsilon n = nV_{\rm{template}}/V$.

How does the covering fraction increase when $n$ becomes larger?  To
understand this, it is helpful to first consider the behavior that the
covering fraction would have in the case where the $n$ points in the
template bank were simply selected at random from $\cal M$, without
any consideration of whether or not they were closer together than
$\Delta$.  This case is considered in some detail in a recent paper on
{\bf random} template banks \cite{Messenger:2008ta}.  (In contrast, this paper uses the name
{\em stochastic} template bank.) In that case, since on
average each additional template removes a fraction $\epsilon$ of the
volume that is not already covered, one obtains
\begin{equation}
E(f(n)) = 1 - \exp(-\epsilon n)
\end{equation}
or
\begin{equation}
E(f(\Theta)) = 1 - \exp(-\Theta)
\end{equation}
for the expectation value of the coverage.
For small $n$, this gives a linear increase in the covering fraction,
which also describes the stochastic template bank.

\begin{figure}
\includegraphics[width=3in,angle=0]{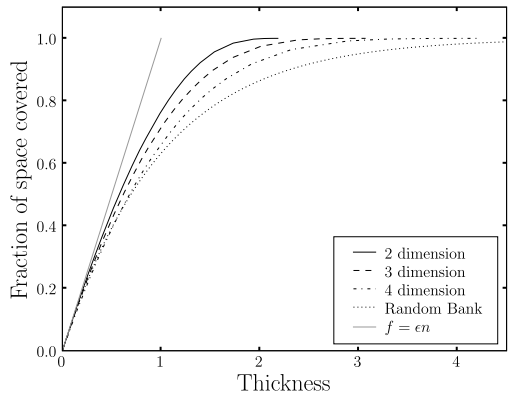}
\caption{The relationship between the 
covering fraction and the thickness
of the bank in 2, 3 and 4 dimensions.
This is also compared to what one would expect
in the case of the random template bank \cite{Messenger:2008ta},
as well as the case where no templates overlap each other.
\label{fig:section2_comp}
}
\end{figure}

Compared to the random template bank, on the average, a stochastic
template bank gives higher coverage for a given number of templates.
This is illustrated in Figure~\ref{fig:section2_comp},
which shows the
covering fraction as a function of thickness, where the
signal manifold $\cal M$ is a unit box in 2, 3 and 4 dimensions.
Thus, if it is desirable to minimize the computation cost because a
single template bank is going to be used and re-used many times, the
stochastic banks could offer a significant improvement compared with the
random ones. The graph does seem to indicate, however, that the stochastic
bank converges toward the random case as dimension increases. Further
investigation is needed to demonstrate what level of improvement the stochastic
bank would have over the random bank at high dimension.

\subsection{Computational cost of filtering templates}

While the stochastic bank will provide a better coverage for the same number of
templates, one must incur an extra computational cost to carry out
the filtering stage of the stochastic placement algorithm. This sub-section
investigates what this computational cost would be as a function of number of
templates and covering fraction.

\begin{figure}
\includegraphics[width=3in,angle=0]{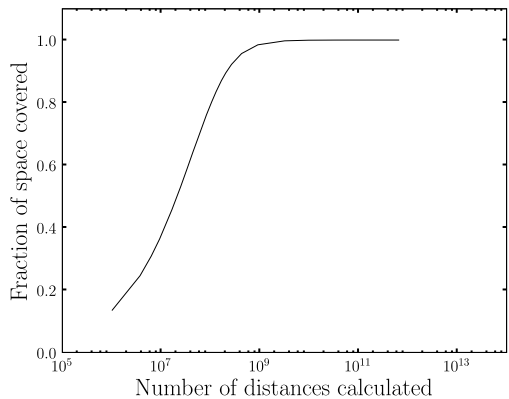}
\caption{The computational cost (number of
distance calculations) depends on the covering fraction.
\label{fig:section2_cost} }
\end{figure}

If every random
point was accepted as a template, because it was farther than $\Delta$
from all previous templates, then the computational cost would be
\begin{equation}
C = \alpha n (n-1)/2,
\end{equation}
where $\alpha$ is the cost of computing the distance between two
points. This follows because the distance must be calculated between
all possible pairs of templates, and there are $n(n-1)/2$ such pairs.
This also correctly describes the cost of stochastic template bank
creation when the covering fraction is substantially less than one,
and few potential templates are rejected.  But when the covering
fraction approaches one, the computational cost explodes, because the
dominant computational cost is the cost of rejecting templates.
This is shown in Figure~\ref{fig:section2_cost}.

This also allows us to provide an estimate of the computational cost.
In practice, 100\% coverage is not necessary or desired. For a typical binary
inspiral search one might be happy with a coverage $f \in [0.9, 0.99]$.  For such coverages
the computation cost is bounded above by
\begin{equation}
C = \frac{\alpha}{1-f} \frac{n_{\rm estimated}^2}{2},
\end{equation}
which is obtained by assuming that the cost of adding the last template is the
same as the cost of adding every template.
This is an upper bound because the factor of $1/2$ is larger than
$n(f=1/2)/ n_{\rm estimated}$, and because the computational cost of
adding the earlier templates is smaller than that of adding the final
template.

The computational cost of this method grows faster than the square of
the number of templates.  However, there is a modified version of this
algorithm in development that has a cost proportional to $n \log n$.  This maintains
an internal list of hyper-cubic cells, which contain points separated by distances
of less than $2\Delta$.  When a new random template is considered, its
distance only needs to be compared to the points in the same cell, and
the $3^D-1$ neighboring cells.  The process of looking to see if there
are neighboring cells requires a binary search in an index list, and
accounts for the additional $\log n$ factor.

It is this prohibitive computational cost that has prevented us from being able
to test the stochastic template bank in dimensions higher than 4 without boundary
effects becoming rather pronounced. With this improved version of the algorithm
it is hoped that a test of the stochastic bank in higher dimensions can be
performed.

\section{Testing the algorithm}
\label{sec:tests}

This section investigates how the stochastic template placement performs in
dimensions less than 5 and how this compares with geometrical placement algorithms.
An investigation of how the algorithm performs when the distribution of initial
seed points is not proportional to the determinant of the metric is carried out as well
as a demonstration that the stochastic algorithm will perform well in intrinsically
curved parameter spaces.

\subsection{Templates in flat spaces of different dimensions}
\label{sec:dimensions}
First consider a flat unit hyper-cube in $D$-dimensions, with Cartesian
coordinates and the metric $g_{ij}=\delta_{ij}.$ Each coordinate
lies in the range $[0,1]$.  $\Delta$ is chosen so that $1/\epsilon$
is equal to 10000.

\begin{figure}
\includegraphics[width=3in,angle=0]{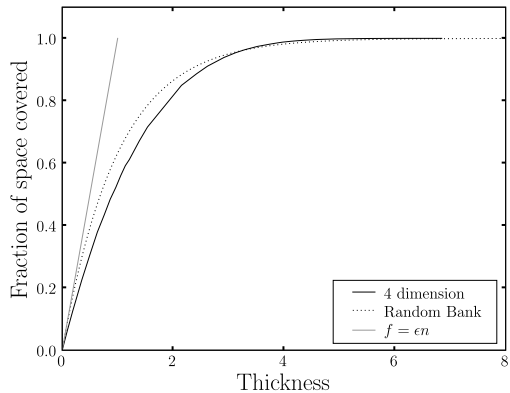}
\caption{As Figure \ref{fig:section2_comp}
but setting the value of $\rm{N_{lower-bound}}$ to be 50. By comparing the two figures one 
can see how boundary effects manifest themselves both by decreasing the initial
slope $df/dn$ and by requiring a much larger number of templates than the estimate.
\label{fig:smaller_comp} }
\end{figure}

Figures \ref{fig:section2_comp} and \ref{fig:section2_cost} show the coverage,
in 2, 3 and 4 dimension, and computational cost
as a function of the number of templates in the bank.
The coverage in Fig.\,\ref{fig:section2_comp} was computed using
Monte-Carlo integration with 20,000 sample points.
The coverage is the fraction of these points that
are less than $\Delta$ from a template in the bank. To generate Fig.\,\ref{fig:section2_comp}
as well as Figures \ref{fig:smaller_comp}, \ref{fig:section2_cost}, \ref{fig:stochligo01} and Tables
\ref{tab:polar} and \ref{tab:sphere} this process was carried out 100 times and the mean of the
values obtained was used.

Figure \ref{fig:comparison} compares the number of templates being converged upon
by the stochastic bank with the estimates and the lattice algorithms
as described in section \ref{sec:algorithm}.
It can be seen from this table that the stochastic template banks
perform better than the naive hyper-cubic lattices as the dimension $D$ of
the parameter space increases.  This is what was predicted in the
previous section: the stochastic template banks converge to
``complete'' coverage with fewer templates than would be needed in a
cubic lattice. Also, as predicted, the $A_n^*$ lattice is more efficient
than the stochastic bank when the stochastic bank has reached complete
coverage.

An interesting feature, which is more noticeable when the number of templates
in the banks is reduced,
is that they show effects due
to the boundaries, especially noticeable in the higher dimensions.
This
is easy to understand.  Any template located closer than distance
$\Delta$ to the boundary of the unit $D-$cube will have part of its
coverage region lying outside the cube.  Consequently, if $\Delta$ is
too large, then many of the templates will fail to produce the amount
of coverage that would arise if no boundaries were present.  Thus, a
sign that boundary effects are appearing is that the initial coverage grows
more slowly with template number than expected.

This effect can be seen in Figure
\ref{fig:smaller_comp}
where the initial slope $df/d\Theta$ at
$\Theta=0$ is smaller than unity and 
also the final thickness is much larger than the
estimate, which was not seen in Figure \ref{fig:section2_comp}.  At what template
radius $\Delta$ do boundary effects become significant?  This can be
easily understood by estimating the volume that lies within distance
$\Delta$ of the boundary of the $D-$cube.  This is $V_{\Delta{\rm 
-boundary}} = V - (V^{1/D} - 2 \Delta)^{D}  \approx 2D \, \Delta \, V ^{1-1/D}$.  
Hence the initial coverage, when
$\Theta \ll 1 ,$ is
\begin{equation}
\left( \frac{df}{d\Theta} \right)_{\Theta=0} =
1 - \beta \frac{V_{\Delta{\rm -boundary}}}{V}.
\end{equation}
Here $\beta$ is a numerical factor, of order $1/6$ in three
dimensions, which measures the average fraction of volume of a
template that lies outside the cube, as the center of the template
moves through all positions in the $\Delta{\rm -boundary}$.

\subsection{Choice of coordinate system and convergence of template numbers}

How does the convergence of the stochastic template bank generation
depend upon the distribution of the random template candidates in the
underlying parameter space?  This question is of practical interest,
because the optimal distribution of the random points has a
probability proportional to the the volume element
$\sqrt{\det{(g_{ab})}} {\rm d}^Dx$.  However, it can be difficult in practice
to generate such a distribution, whereas it is simple to generate
random points that have a uniform distribution in the coordinates.

\begin{table}
\begin{tabular}{c|cccc|cccc}
\hline
\hline
& \multicolumn{4}{|c|}{Cartesian} & \multicolumn{4}{|c}{Polar} \\
$N$       &  $n$    & $\sigma_{n}$  &  $f$   &  $\sigma_{f}$ 
            &  $n$    & $\sigma_{n}$  &  $f$  &  $\sigma_{f}$\\
\hline
1500       & 1397.1 & 1.0 &  0.1353  & 0.0003 & 1313.4 & 1.3 & 0.1246 & 0.0003\\
5000	  & 3994.4 & 2.6 &  0.3632  & 0.0004 & 3613.9 & 2.6 & 0.3237 & 0.0004\\       
15000      & 8494.5 & 3.9 &  0.6818  & 0.0004 & 7605.3 & 3.9 & 0.6140 & 0.0004\\
50000      & 13961.5 & 4.1 &  0.9221  & 0.0002 & 12979.6 & 4.0 & 0.8825 & 0.0003\\
150000     & 17307.8 & 4.0 &  0.9847  & 0.0001 & 16676.6 & 3.6 & 0.9747 & 0.0001\\
500000     & 19365.5 & 3.1 &  0.99746  & 0.00004 & 19025.4 & 3.5 & 0.99582 & 0.00005\\
1500000    & 20439.3 & 3.0 &  0.99949  & 0.00001 & 20241.3 & 3.6 & 0.99917 & 0.00002\\
5000000    & 21141.9 & 3.5 &  0.99990  & 0.00001 & 21023.4 & 3.2 & 0.99987 & 0.00001\\
10000000   & 21401.3 & 3.1 &  0.999971  & 0.000003 & 21305.4 & 3.3 & 0.999948 & 0.000005\\ 
\hline
\hline
\end{tabular}
\caption{Number of templates $n$ and fractional coverage $f$ with 
associated standard deviations as 
a function of the cumulative number of trials $N$ in the case of 
Cartesian ($n_C$) and polar ($n_P$) coordinates.}
\label{tab:polar}
\end{table}

\begin{figure}
\includegraphics[width=3in,angle=0]{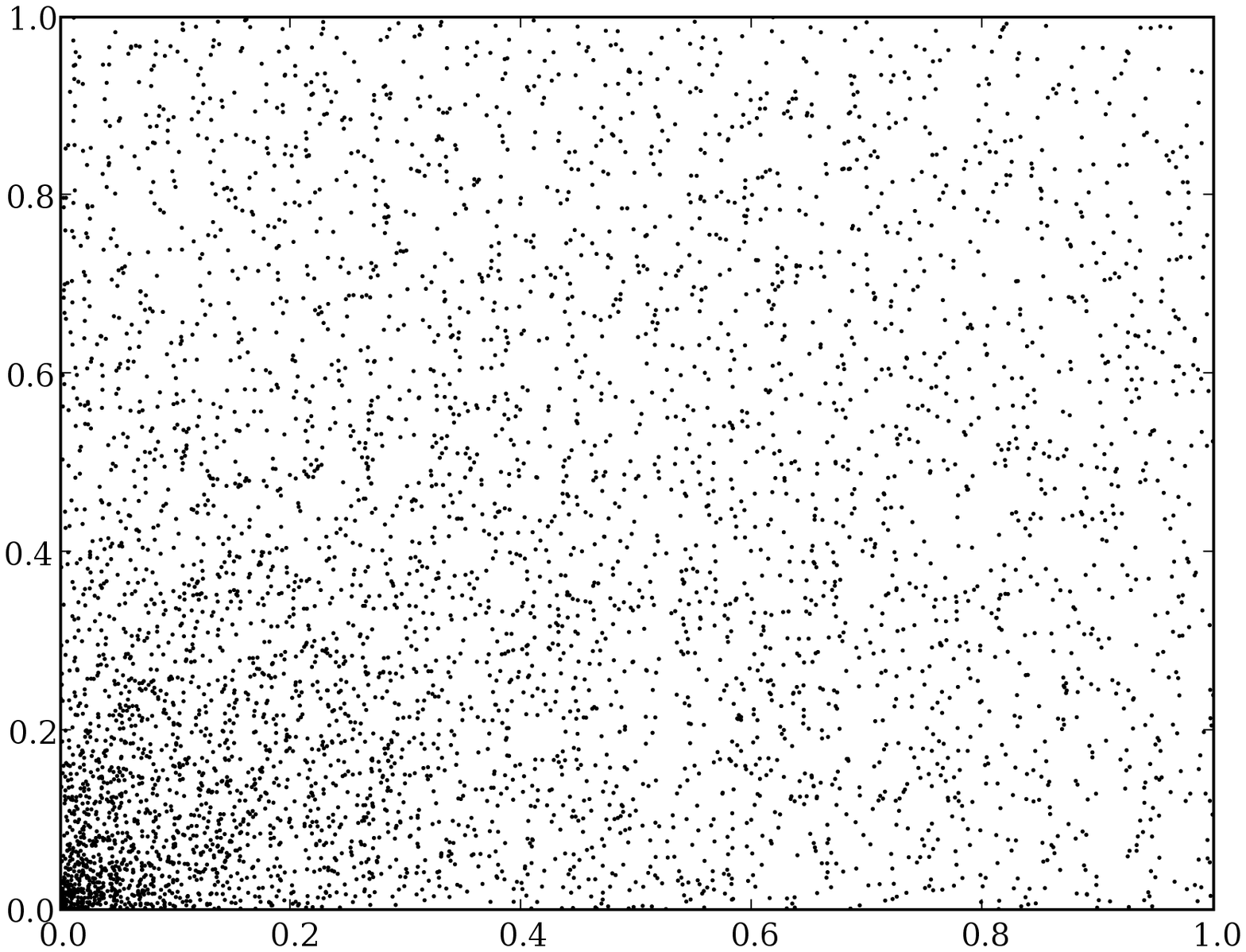}
\includegraphics[width=3in,angle=0]{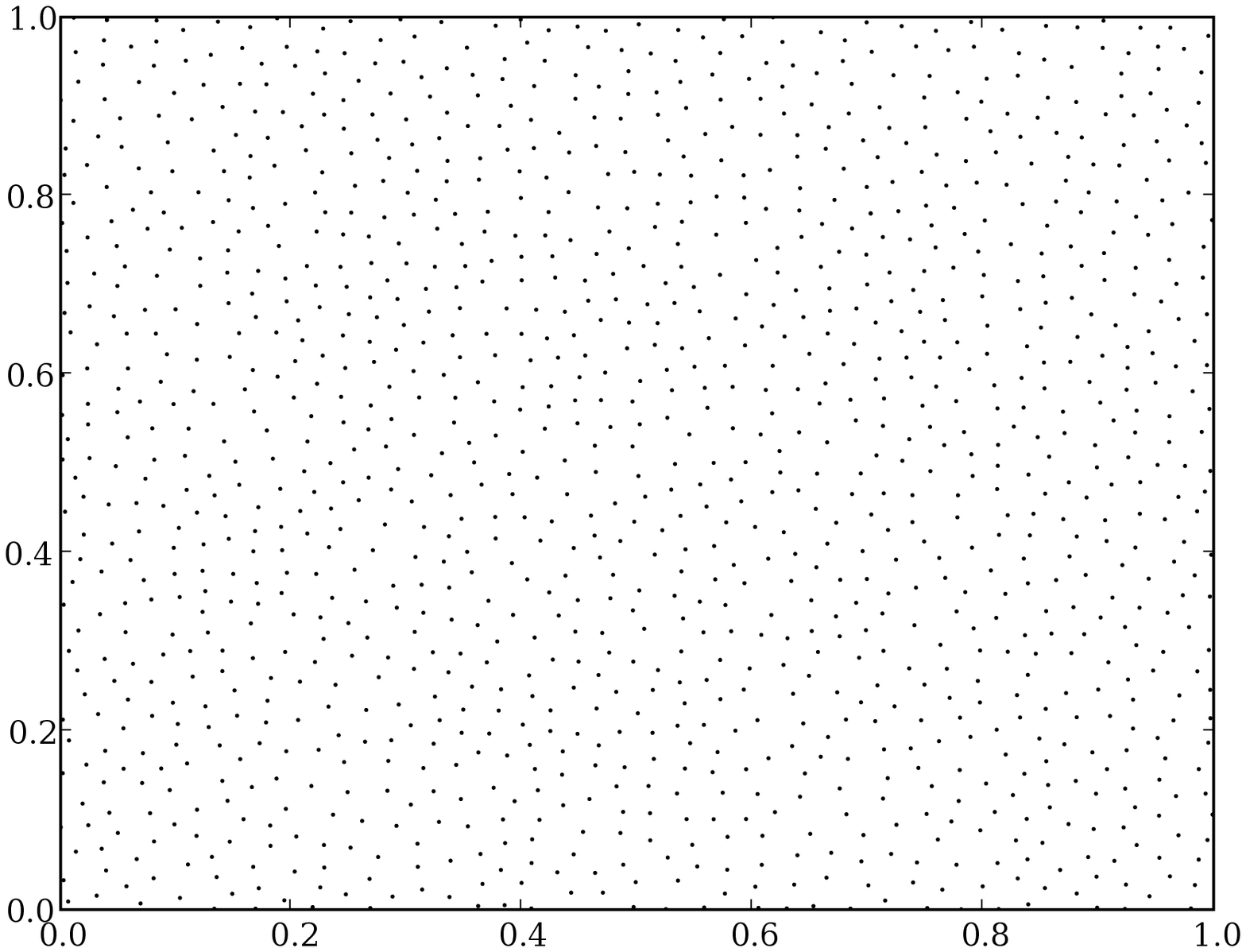}
\caption{The distribution of trial points chosen uniformly in polar
coordinates (left panel) and the points that remain as templates after the
application of the stochastic placement algorithm (right panel).  
\label{fig:polar} }
\end{figure}

For example, in two-dimensional flat space, one could choose trial
points with uniform probability distributions in polar coordinates.
This means that too many random templates are tested
from the region near the origin, and then rejected.  However, they are soon rejected, as
being too close to points already in the template bank, and in the
end, the template points that survive have the correct probability
distribution proportional to $dx\,dy = r\, d\theta\, dr$.  This is shown in 
Figure \ref{fig:polar}.

Table\,\ref{tab:polar} shows the number of templates $n$ as a function of
the number of trial points $N$ for random template candidates distributed uniformly in Cartesian and
polar coordinates. Also shown is the coverage of the template bank, calculated using Monte-Carlo integration as
described in the previous section.

This is very useful because in many cases one does not know
coordinates in which the determinant of the metric is constant.  Of
course one could simply distribute points with a probability density
proportional to the volume measure!

\subsection{Templates on a sphere}

So far the paper has considered templates in a flat signal space. 
However one can consider examples where the signal manifold is not flat, but is
curved.  This introduces two new issues.

First, the distance between widely separated points can no longer be
easily computed. However, the only important case is the one in which
the points are nearby.  In this case, one can use the metric to
approximate the distance at small separations:
\begin{equation}
{\rm d} l^2 = g_{ik}(x^j_A)  \left(x^i_A - x^i_B \right ) \left(x^k_A - x^k_B \right ).
\label{eq:metricapprox}
\end{equation}
Since the components of the metric can be expensive to calculate, an
efficient approach is to calculate and store those components only
for points that are included in the template bank.  Those metric
components are then used for the distance comparisons with potential
new (randomly chosen) template candidates.

Second, depending upon the choice of the coordinate system, the
determinant of the metric may be non-constant.  In this case, 
an efficient approach would be to generate random points with a
probability distribution proportional to the volume element
$\sqrt{\det{(g_{ab})}} {\rm d}^Dx$.  However, in practice one can generate
points with \emph{any} distribution in the coordinates: the stochastic
template placement algorithm simply rejects those points that are not
needed, and produces a distribution with the correct density
proportional to $\sqrt{\det{(g_{ab})}} {\rm d}^Dx$.

To demonstrate the performance of the stochastic template placement
algorithm on a curved manifold, consider a unit-radius two-sphere $S^2$ with
standard spherical polar coordinates $(\theta, \varphi)$.  The metric
is
\begin{equation}
{\rm d}l^2 = {\rm d}\theta^2 + \sin^2\theta\, {\rm d}\varphi^2.
\end{equation}
Table\,\ref{tab:sphere} shows the number of templates $n$ as a function
of the number of trial points $N.$
The size of the templates has been chosen so that the
ratio $ \epsilon =V_{\rm template}/V$ is the same as for the unit cube
examples given in the previous section.  In this case the stochastic
algorithm converges for a smaller number of templates than for the unit
cube.  This is for the reasons described above: since the unit sphere
has no boundary, no templates lie partly outside the space, so every
template provides the largest possible coverage.

\begin{figure}
\includegraphics[width=0.32\linewidth,angle=0]{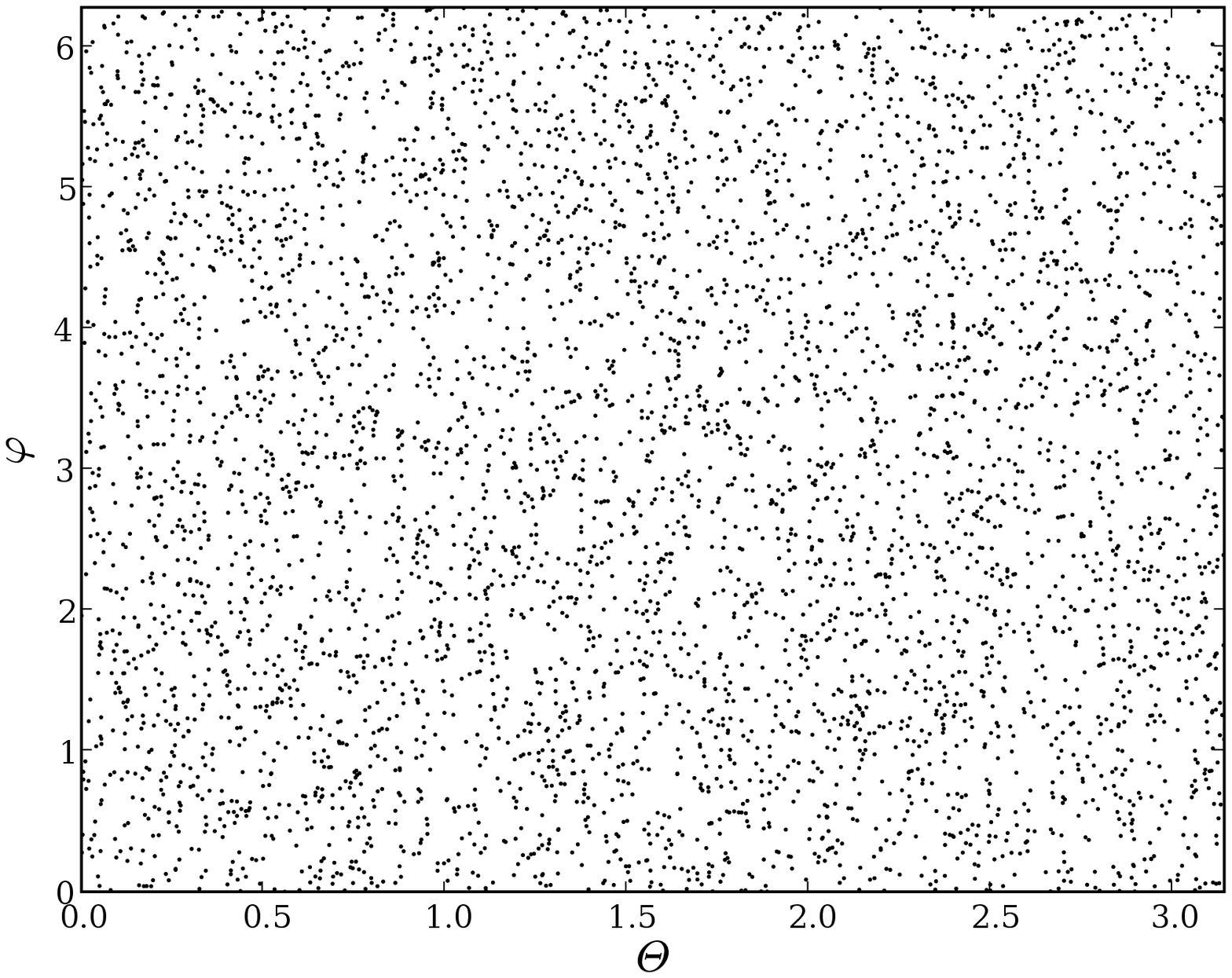}
\includegraphics[width=0.32\linewidth,angle=0]{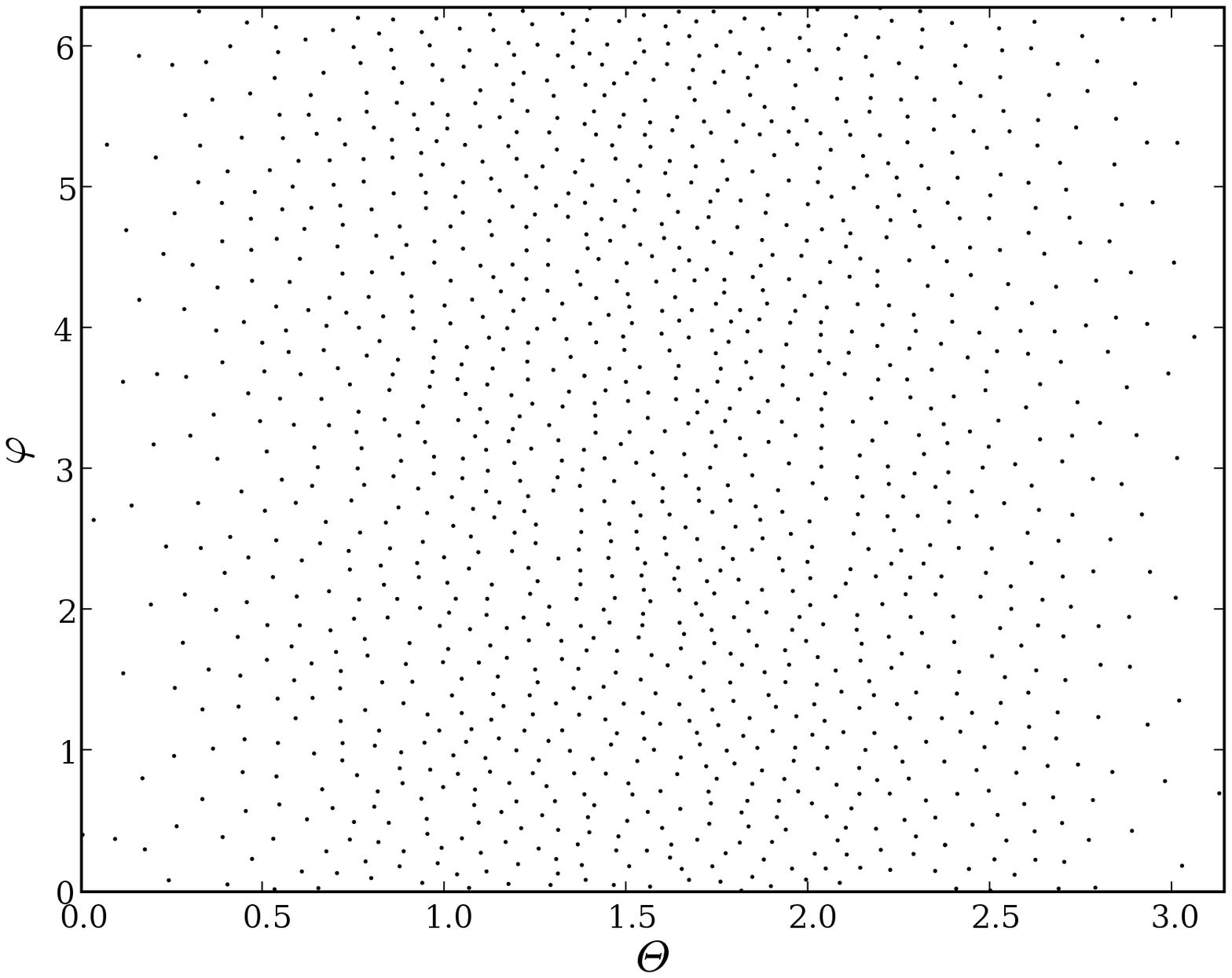}
\includegraphics[width=0.32\linewidth,angle=0]{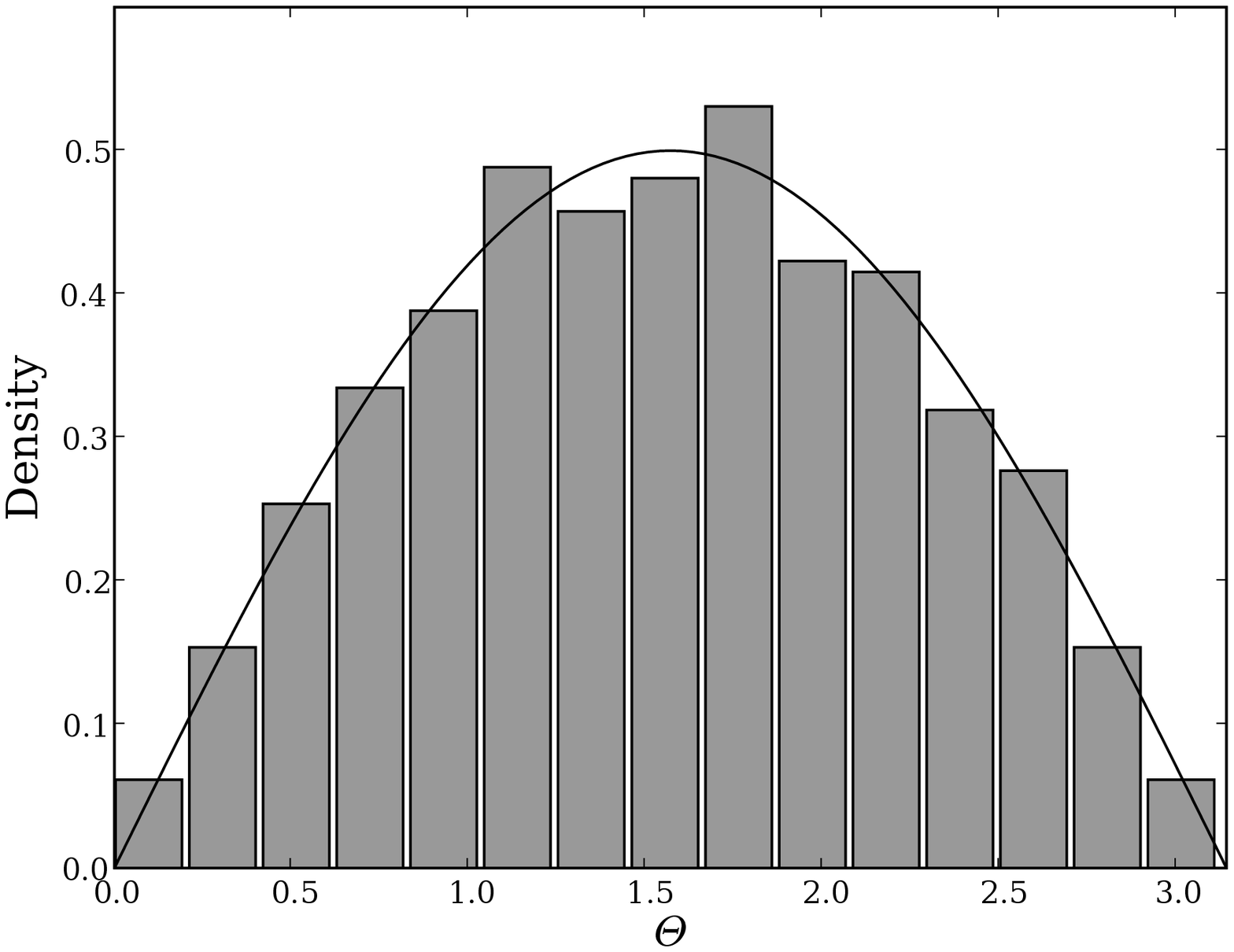}
\caption{Trial points chosen uniformly in the $(\theta,\,\varphi)$ 
coordinates (left panel) and the templates on the surface of a
sphere of unit area
that remain after the application of the stochastic template placement algorithm
(middle panel). Also the distribution of these templates in
$\theta$-coordinate (right panel)
where the solid line shows the expected distribution.
\label{fig:sphere} }
\end{figure}

Fig.\,\ref{fig:sphere} shows the distribution of 5000
candidate points, chosen uniformly in spherical polar coordinates
$(\theta,\,\varphi)$ (top panel).  The points that survive and remain
in the stochastic template bank are shown in the middle panel. A
histogram of the distribution of the templates as a function of
$\theta$ is also shown (right panel). As expected, the density of
templates is proportional to the volume element $\sin \theta\, {\rm d}\theta\,
{\rm d}\phi$: it is smallest at the poles and the greatest at the equator.

\section{Templates for gravitational wave chirps}
\label{sec:chirps}
Binary systems of compact objects (i.e., black holes and/or neutron stars)
evolve by emitting gravitational radiation. The loss of energy and angular
momentum into gravitational waves causes the two bodies to spiral in
toward each other, emitting a burst of radiation just before they
merge. Although there is no exact solution to the two-body problem in
general relativity, an approximation method called the post-Newtonian
formalism has been used to compute the amplitude and phase of the 
waves emitted in the adiabatic inspiral phase  to a very high accuracy
\cite{BFIJ02,BDEI04,ABIQ04} (for a recent review see Ref.\,\cite{Bliving}).
Moreover, recent progress in numerical relativity
has provided a good knowledge of the waveform even in the strong
gravity regime of the merger dynamics \cite{Pretorius05}. Thus, one can use matched filtering
to dig out astrophysical signals from the noise background of an interferometric
detector. 

In general, the radiation from a binary is characterized by as many as
seventeen parameters. However, some of these parameters  (the distance to the binary, the 
inclination of the orbit relative to the detector, etc.)
only affect the
amplitude of the waveform, 
which does not modify the search template.
Therefore, one would only need to place
templates in a lower-dimensional parameter space (say six or seven dimensions).
State-of-the-art template placement algorithms deal only with a binary 
composed of non-spinning objects (in which case templates are only needed
in the two-dimensional parameter space of the masses of the component stars)
or at best a simplified model of a binary composed of spinning objects
requiring one, or two, additional dimensions. Clearly, this is not satisfactory
as there is no reason to believe that an astronomical binary will respect
simplified models.

The goal of the stochastic template placement algorithm
is to address the problem of choosing templates on a manifold of arbitrary
dimensions. However, in this paper, the algorithm is only applied
to the case of a binary consisting of non-spinning bodies
where the results are well known, thus facilitating a straightforward comparison
with established results. This algorithm has also been applied in a recent
search for spinning binaries in the first year of LIGO's 
fifth science run using templates placed in a three-dimensional 
parameter space \cite{Vandenbroeck:2009cv} as well as a
five-dimensional search for super-massive black holes in a mock data challenge
\cite{IHSFBSS-09}. A similar, but independently developed, algorithm was also
used in this mock data challenge \cite{Babak:2008rb}. This algorithm was
effectively the same as the one described in this work but the author
calculates the overlap between points explicitly, instead of using the metric
approximation as in this work. While this will more accurately determine the
overlap, especially for overlaps not close to unity, it will come at considerable
additional computational cost.

\begin{table}[t]
\begin{tabular}{c|cccc}
\hline
\hline
$N$     &  $n$    & $\sigma_{n}$  &  $f$   &  $\sigma_{f}$  \\
\hline
1500.0      \,\, & 1330.3  & 1.2 & 0.1989 & 0.0004 \\
5000.0      \,\, & 3688.3  & 2.8 & 0.4253 & 0.0004 \\
15000.0     \,\, & 7807.1  & 3.7 & 0.7027  & 0.0004 \\
50000.0     \,\, & 13198.7 & 4.1 & 0.9192 & 0.0002 \\
150000.0    \,\, & 16779.5 & 4.4 & 0.9836 & 0.0001 \\
500000.0    \,\, & 19007.0 & 3.6 & 0.99735 & 0.00003 \\
1500000.0   \,\, & 20171.6 & 3.6 & 0.99947 & 0.00002 \\
5000000.0   \,\, & 20906.6 & 3.8 & 0.99991 & 0.00001 \\
10000000.0  \,\, & 21182.9 & 4.0 & 0.999967 & 0.000004 \\
\hline
\hline
\end{tabular}
\caption{Number of templates $n$ and coverage $f$ as a function of the number of trials $N$ on a sphere of unit radius.}
\label{tab:sphere}
\end{table}

\subsection{Choice of coordinate system}
Begin by choosing a suitable coordinate system on the signal manifold.
The masses $m_1$ and $m_2$ of the component stars are the most obvious 
coordinates on the manifold. However, when one uses masses as the coordinate
system the determinant of the metric will vary significantly over this
parameter space \cite{OwenSathyaprakash98}. Because of this
a much higher density of templates is needed
in the low mass region than in the high mass region.

\begin{figure}[t]
\includegraphics[width=3in,angle=0]{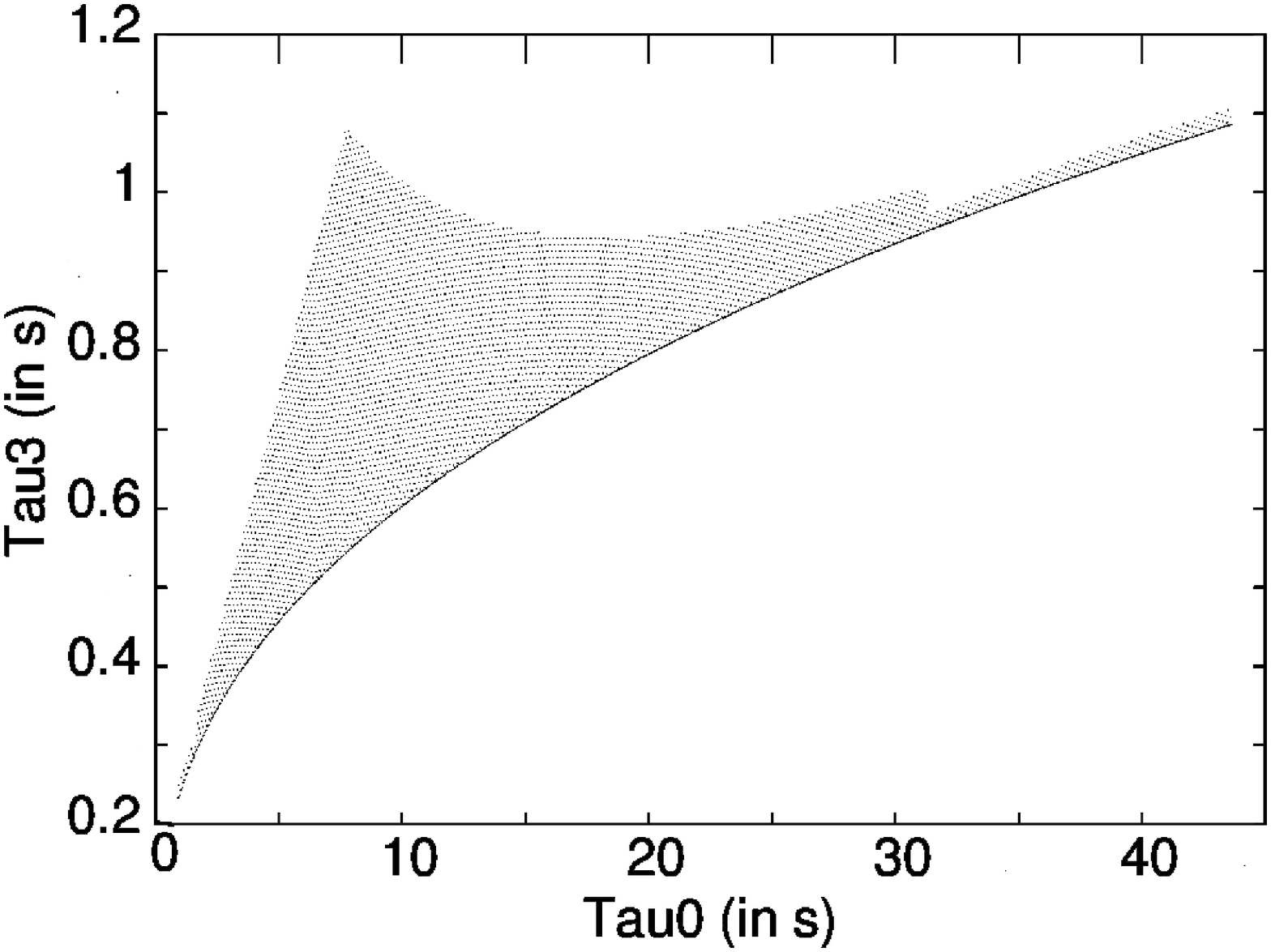}
\includegraphics[width=3in,angle=0]{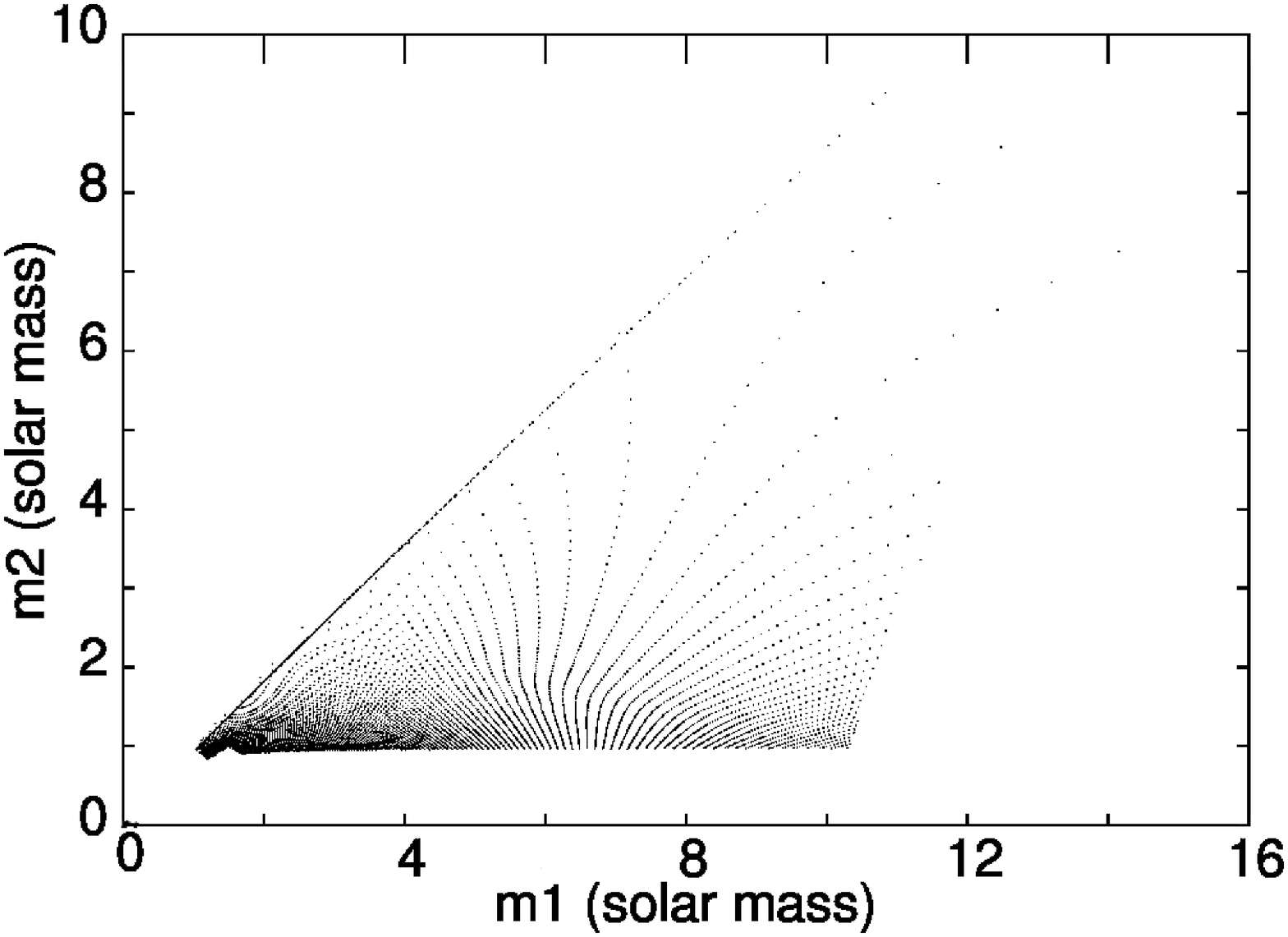}
\caption{The distribution of templates placed by the hexagonal
lattice algorithm in $(\tau_0,\,\tau_3)$
coordinates (left panel) and the same templates in $(m_1,\,m_2)$ 
coordinates (right panel). Clearly, the distribution is highly 
skewed in the latter coordinates.
\label{fig:templates-square-lattice} }
\end{figure}

A better coordinate system is {\it chirp times}
\cite{Sathyaprakash:1994nj}, defined by
\begin{equation}
\theta_1 = \frac{5}{128 \nu} \left(\pi  M f_{\rm L}\right)^{-5/3},\ \ \ \ 
\theta_2 = \frac{-\pi }{4 \nu}  \left( \pi  M  f_{\rm L}\right)^{-2/3}
\label{eq:theta parameters}
\end{equation}
\begin{equation}
\tau_0 = \frac{\theta_1}{2\pi f_{\rm L}},\,\,\,\,
\tau_3 = \frac{\theta_2}{2\pi f_{\rm L}}.
\end{equation}
Using this coordinate system the
determinant of the metric does not vary much over this parameter
space.
This can be illustrated by looking at the distribution of templates
in both coordinate-systems as shown 
in Figure \ref{fig:templates-square-lattice}.
The algorithm used in this case \cite{Cokelaer:2007kx} places templates first
along the $m_1=m_2$ curve (the lower right boundary in the left panel and upper
left boundary in the right panel). This is dictated by the fact that the region
below the equal-masses curve (in $\tau_0$,$\tau_3$ coordinates) corresponds
to binaries with imaginary component masses\footnote{Although the waveform, 
which depends only on the total mass $M$ and mass ratio $\nu$ which are real in that
region, can be generated in this part of the parameter space, it is
unphysical and, therefore, not of any interest.}.  The algorithm uses
a hexagonal placement over the rest of the parameter space.

\subsection{Comparison of stochastic lattice with a square lattice}

Let us now compare the stochastically generated 
template bank with
a hexagonal lattice and with a square lattice. 
In this case the template banks are created to cover binary compact objects
whose components have a mass range of 1 to 10 solar masses such that any real
signal within this range of masses would have an ``overlap'' 
greater than 0.96 with
at least one of the templates in the bank. This overlap, defined by
$1 - \Delta^2$, is calculated using the assumption made
in equation~(\ref{eq:metricapprox}) and the metric defined in \cite{Bank06}.
For the stochastic algorithm the trial points are placed uniformly in 
$(\tau_0,\,\tau_3)$ coordinates (and limited by the restrictions on the masses).
This is also compared to trial points placed uniformly in $(m_1,\,m_2)$
coordinates.

\begin{figure}
\includegraphics[width=3in,angle=0]{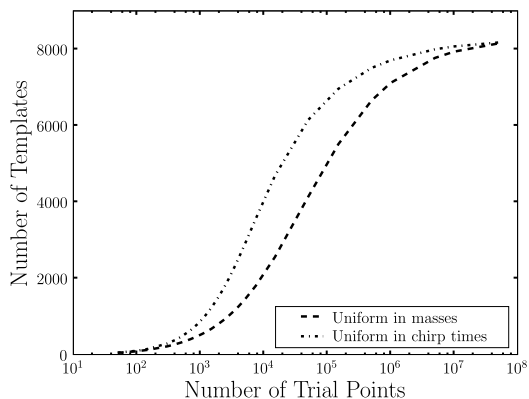}
\caption{The number of templates as a function of the number of random trial points 
is shown when the trial points are assigned uniformly in $(\tau_0,\,\tau_3).$ 
coordinates as well as uniformly in  $(m_1,\,m_2)$ coordinates.  
\label{fig:stochligo01} }
\end{figure}

\begin{figure}
\includegraphics[width=3in,angle=0]{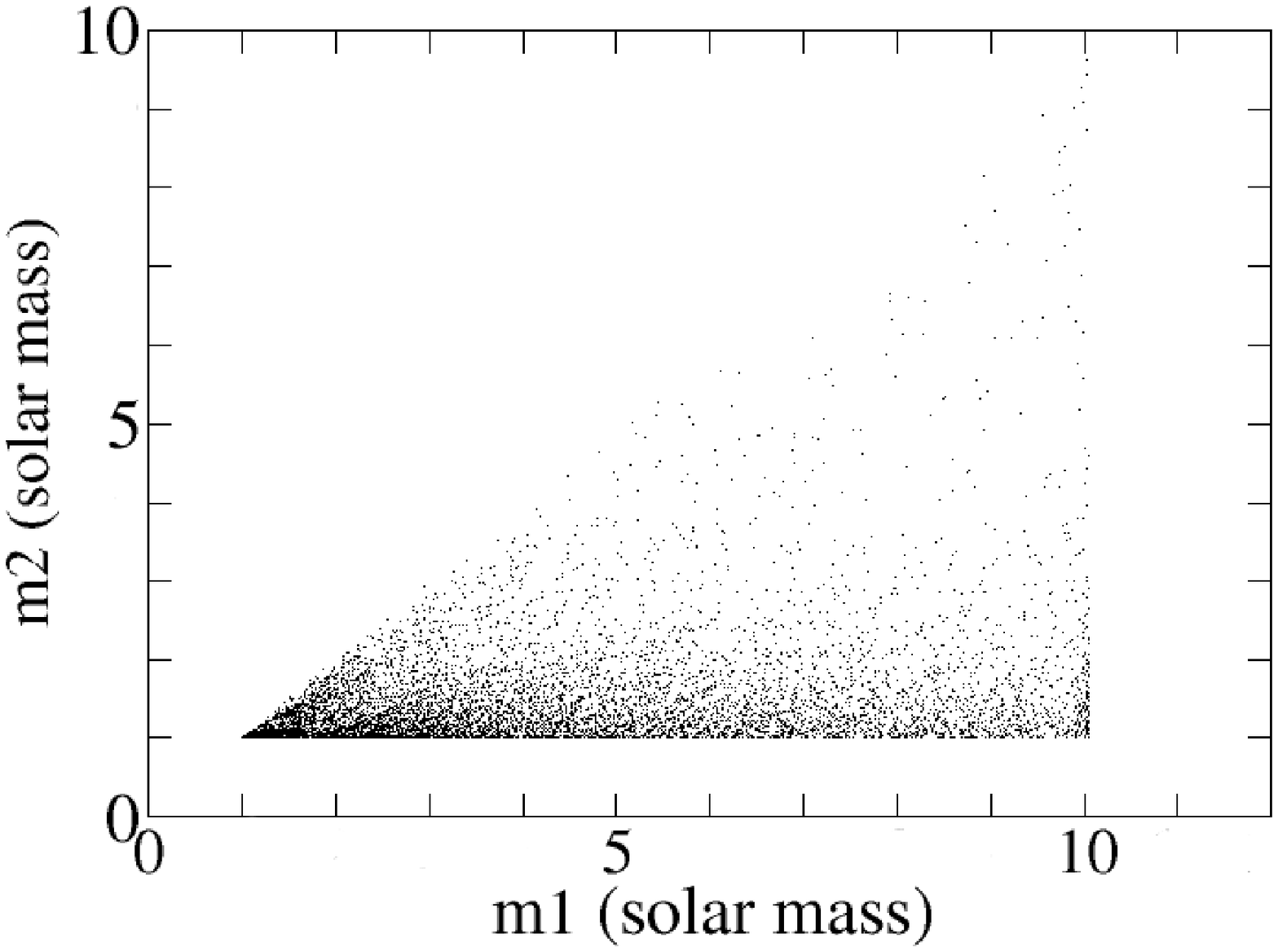}
\includegraphics[width=3in,angle=0]{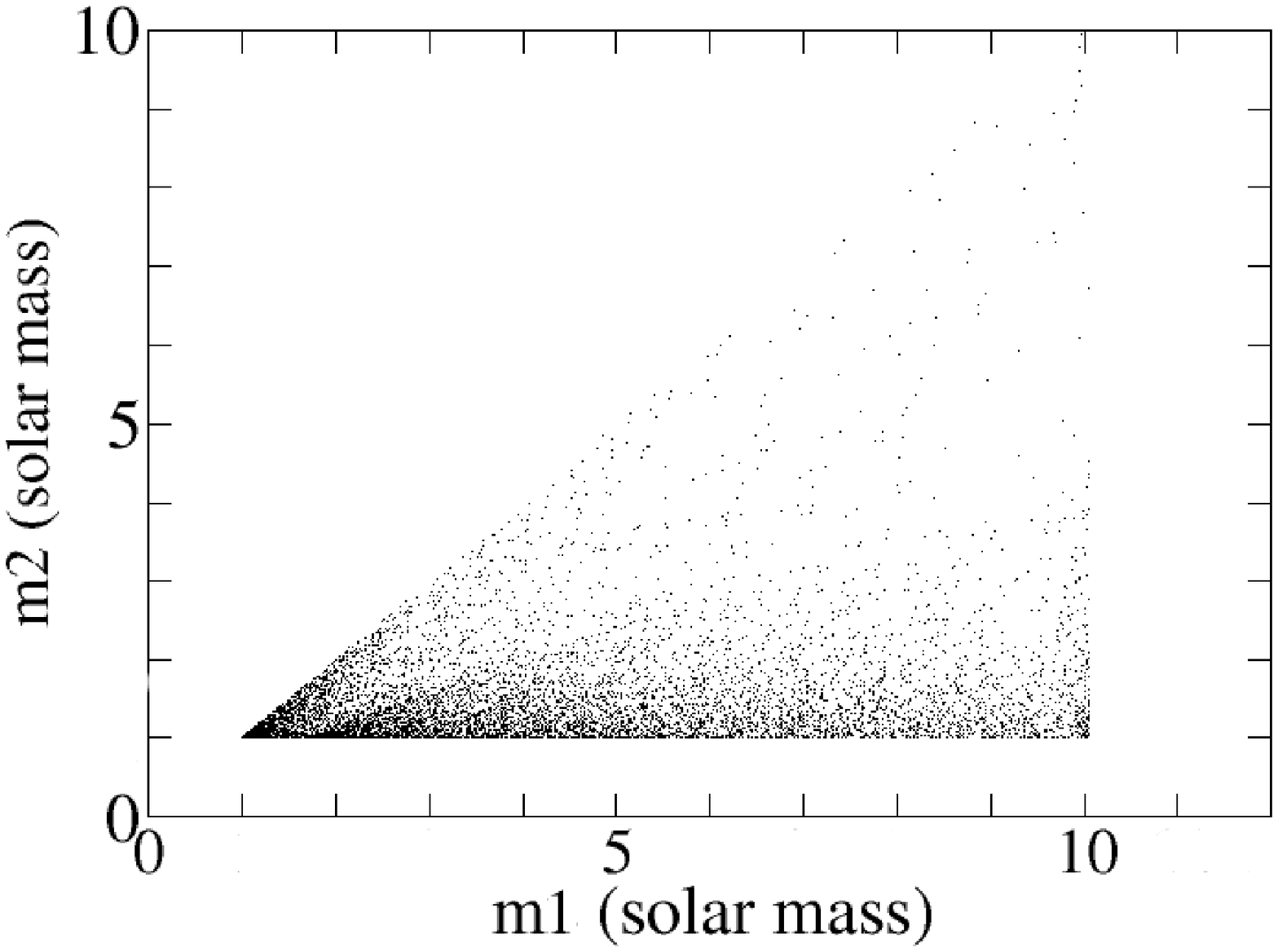}
\caption{The distribution of stochastically generated templates in
$(\tau_0,\,\tau_3).$ coordinates (left panel) and in 
$(m_1,\,m_2)$ coordinates (right panel).
\label{fig:stochligo02} }
\end{figure}

For this choice of parameters and for trial points placed uniformly 
in both coordinate systems, the number of
templates is plotted as a function of the number of trial points
in Fig.\,\ref{fig:stochligo01}.
Fig.\,\ref{fig:stochligo02} shows the distribution of
resultant templates for both initial trial point distributions.

In this two-dimensional example, the stochastic algorithm, in both cases,
converges at about 7500 templates.
For comparison, with the same range of masses 
a hexagonal lattice has 5914 templates and 
and a square lattice has 8353 templates. This may seem to be in conflict with
the statement in section \ref{sec:algorithm} that the stochastic algorithm
performs worse than hyper-cubic lattices in two dimensions. However, one
must remember that the geometrical algorithm used here begins by placing
templates along the boundaries, which is quite inefficient. One also must
remember that though this parameter space is close to flat it is not flat.

\subsection{Efficiency of the Stochastic bank}

The quality or performance of a template bank can be 
assessed by measuring the overlap between randomly simulated
compact binary signals in the relevant range of parameters 
of the template bank in question. To test the performance of the
stochastic template banks,
a set of 20,000 signals (standard post-Newtonian
waveforms of type TaylorT3 \cite{DIS01}) was generated
and the maximum overlap of each over the entire
template bank was calculated. In this case templates
with masses between 3 and 30 solar masses were used 
(a different mass range was used
here to produce less templates, thus making it easier to
show the results graphically) and an
overlap of 0.95 was used, equivalently, $\Delta^2=0.05$.
The template bank was generated from 60,000 trial
points placed uniformly in $(\tau_0,\,\tau_3)$ 
coordinates. 

\begin{figure}
\includegraphics[width=0.32\linewidth,angle=0]{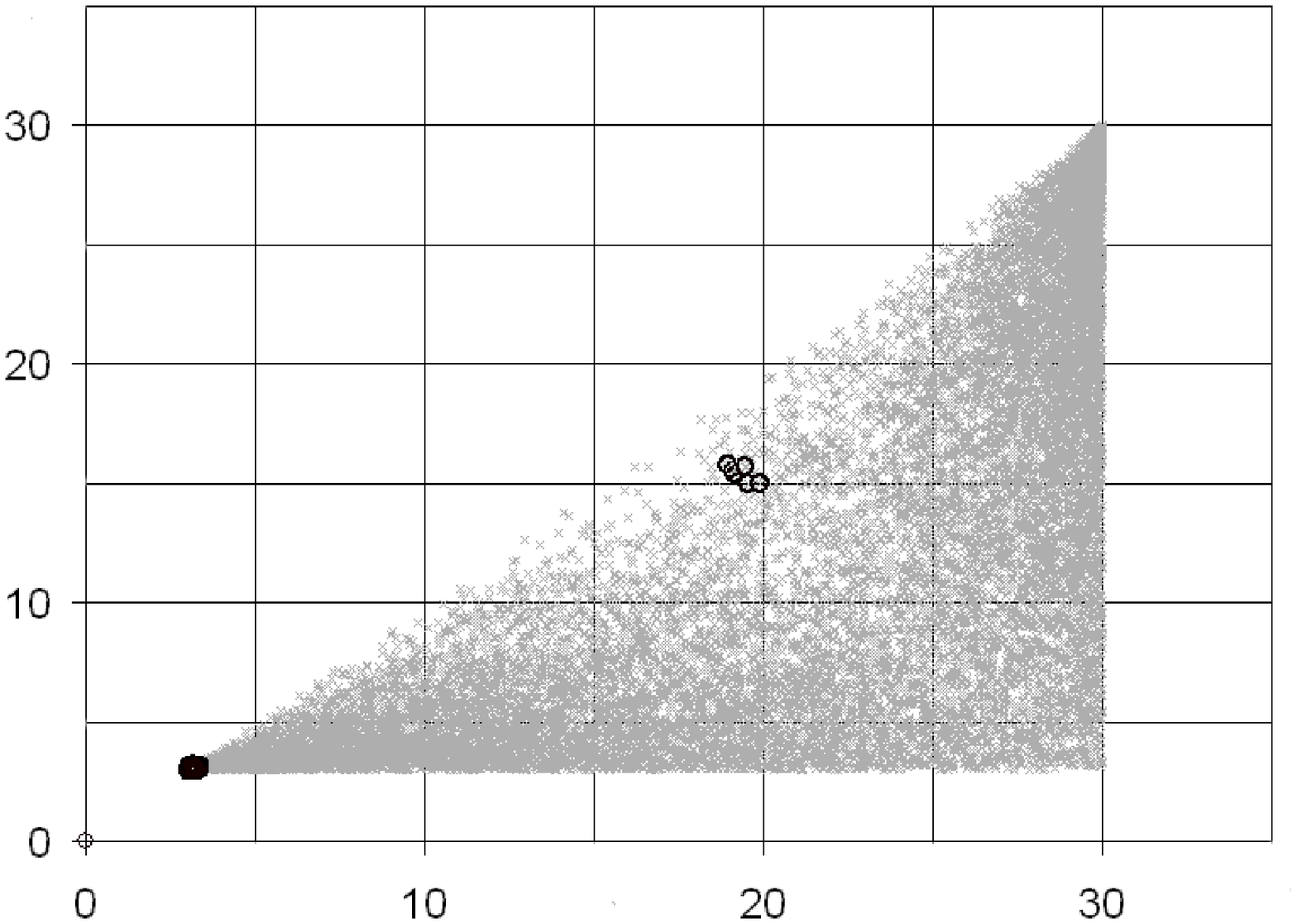}
\includegraphics[width=0.32\linewidth,angle=0]{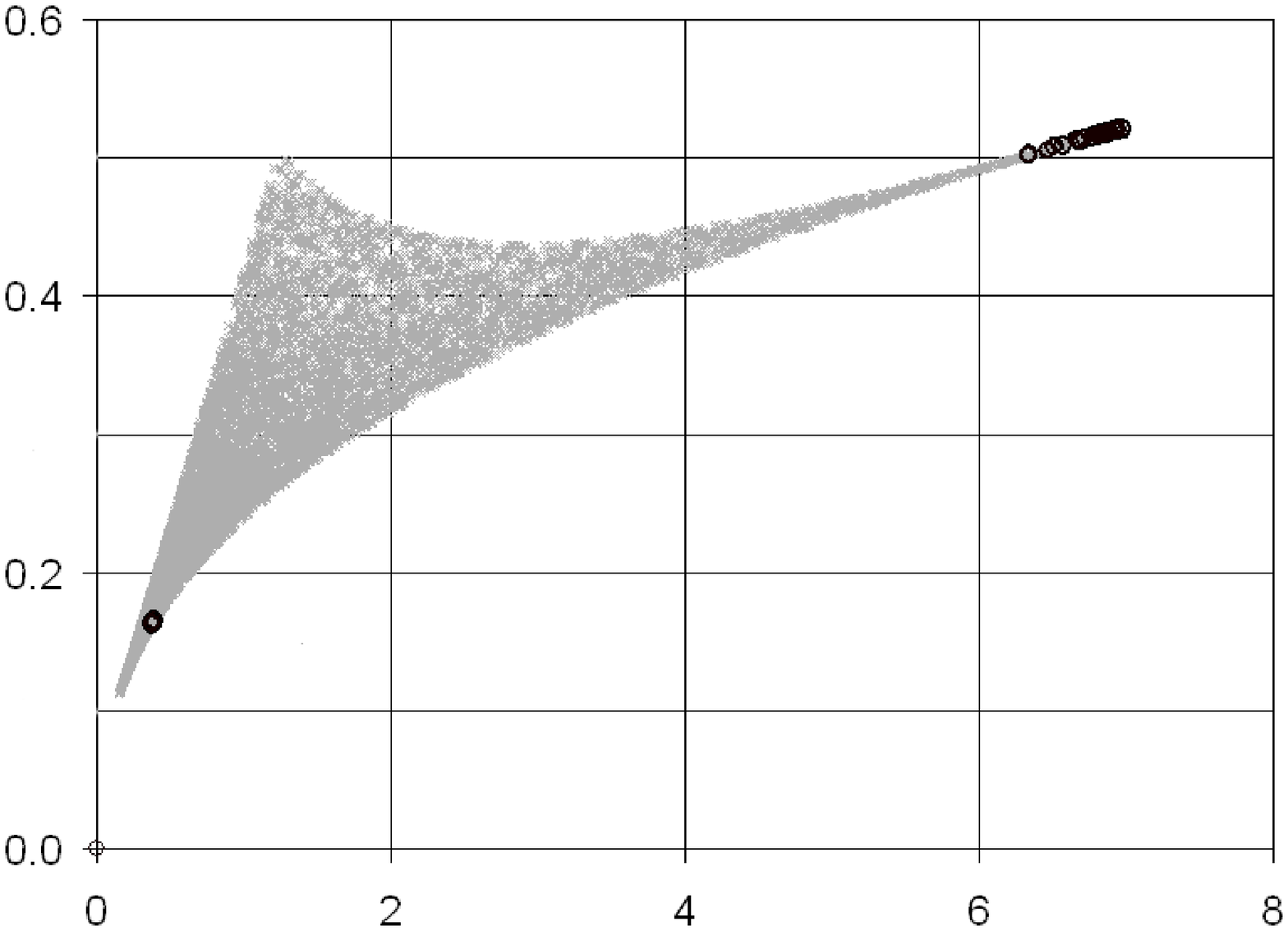}
\includegraphics[width=0.32\linewidth,angle=0]{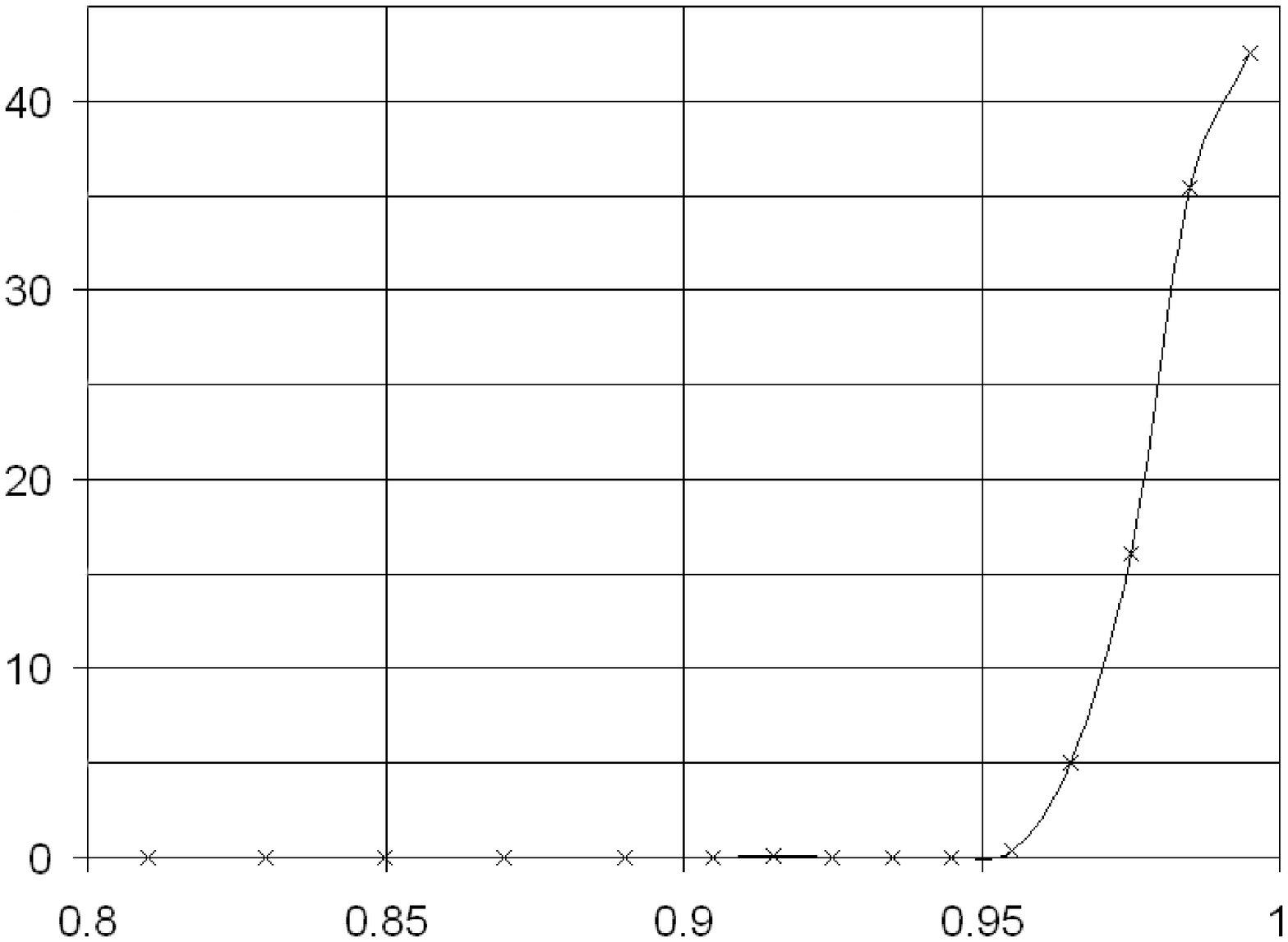}
\caption{The signals which had an overlap larger than 0.95 (gray
crosses) as well as signals with an overlap less than 0.95 (black
circles) in $(m_1,\,m_2)$ coordinates (left) and $(\tau_0,\,\tau_3)$
coordinates (middle). Also a histogram of the overlaps of all injections with
the stochastic template bank (right).
\label{fig:trb01} }
\end{figure}

The result of the test is shown in Fig.\,\ref{fig:trb01}.  One can see
that the stochastic placement algorithm struggles to cover certain
areas of this parameter space. If a larger
number of trial points had been used, the coverage would have been
better.

The areas of the parameter space with poor coverage from
the stochastic template bank are in regions
of the parameter space that are very thin, almost one-dimensional. 
The hexagonal and square lattices also have
this difficulty but they have been specifically designed to 
overcome this problem by placing templates 
along the boundary of the space, especially along the $m_1=m_2$
curve. A stochastic placement algorithm can overcome this problem
in the same manner, or 
by increasing the mass range of allowed templates.  But both solutions come with the
cost of additional templates in the bank. 

\section{Conclusions}

This paper presents a method for stochastically generating template
banks in parameter spaces of arbitrary dimension and with arbitrary
metrics. The relationship between coverage and the number of templates
required to reach that coverage has been investigated for dimensions up
to 4. The performance of the stochastic placement algorithm has been
compared to lattice placement algorithms in flat spaces and was found to
only be marginally less effective at dimension less than 4.
The area where we believe this algorithm would be of most use is in
signal manifolds that have a
large intrinsic curvature, where lattice placement algorithms can not easily be
applied. Stochastic banks which cover less than 100\%
of the signal manifold may be useful for large dimensional manifolds,
though further investigation is needed to show that this is the case.

For cases where the number of required templates is very high, the
algorithm will become very computationally expensive. In these cases
other ``random template banks'', which do not use our filtering stage
might become more practical \cite{Messenger:2008ta}.  Nevertheless the
stochastic template bank will provide better coverage for a given
number of templates.  The construction of stochastic template banks
can be made less expensive, however, by utilizing the fact that it is not necessary to
compute the distance between a trial point and \emph{every} template in the bank.
This is a topic of ongoing investigation.

\section*{Acknowledgements}

We would like to thank Reinhard Prix, Thomas Dent, Stephen Fairhurst, Gareth Jones
and Gian Mario Manca for their
useful comments and feedback in bringing this paper to fruition. In this work
IWH was supported by the Science and Technology Facilities Council, UK,
studentship ST/F005954/1,
BA was partly supported by US National Science 
Foundation grant PHY-0701817 and BSS was supported by Science and Technology
Facilities Council, UK, grant PP/F001096/1.  This paper has been assigned 
LIGO document number P0900069-v3.

\bibliography{references}
\end{document}